\documentclass[prl,aps,twocolumn,amsmath,amssymb]{revtex4}

\usepackage{graphicx}
\usepackage{dcolumn}
\usepackage{bm}

\begin{document}

\preprint{APS/123-QED}

\title{Phase coherence in the inelastic cotunneling regime}

\author{Martin Sigrist,$^1$ Thomas Ihn,$^1$ Klaus Ensslin,$^1$ Daniel Loss,$^2$ Matthias Reinwald,$^3$ and Werner Wegscheider$^3$}
\affiliation{
$^1$Solid State Physics Laboratory, ETH Z\"urich, 8093 Z\"urich, Switzerland\\
$^2$ Department of Physics and Astronomy, University of Basel, Klingelbergstrasse 82, CH-4056 Basel, Switzerland\\
$^3$Institut f\"ur experimentelle und angewandte Physik, Universit\"at Regensburg, Germany
}

\date{\today}

\begin{abstract}
Two quantum dots with tunable mutual tunnel coupling have been
embedded in a two-terminal Aharonov-Bohm geometry.
Aharonov-Bohm oscillations are investigated in the cotunneling
regime. Visibilities of more than 0.8 are measured indicating that
phase-coherent processes are involved in
the elastic and inelastic cotunneling.
An oscillation-phase change of $\pi$ is detected as a function of
bias voltage at the inelastic cotunneling onset.
\end{abstract}

\pacs{Valid PACS appear here}
\maketitle


Is electron transport through quantum dots phase-coherent? This
question roots in the discussion of how to describe it: by
incoherent sequential tunneling, or by coherent resonant
tunneling? A few experiments have shown through the observation of
interference effects that
the current through quantum dots (QDs) has phase coherent
contributions~\cite{Yacoby1995,Goldhaber1998,Holleitner2001,Kobayashi2002}.
In pioneering experiments a
single QD was embedded in an Aharonov-Bohm (AB)
interferometer and AB oscillations were detected on
conductance resonances of the dot~\cite{Yacoby1995}. A QD
molecule with source and drain contacts common to both dots has been
reported to exhibit AB-oscillations when the tunnel
coupling between the dots is negligible~\cite{Holleitner2001}.
Further evidence for phase-coherent transport through QDs can be
deduced from the observation of the Fano-effect in a ring geometry
~\cite{Kobayashi2002} and from the Kondo
Effect in QDs~\cite{Goldhaber1998}.
The question has attracted even more attention due to
proposals to use QDs as qubits~\cite{Loss1998}.
It has been proposed
that entanglement of singlet and triplet states can be probed by
their distinct AB phases~\cite{Loss2000}. Theoreticians
discuss in how far interactions in QDs dephase the
transmitted electrons~\cite{Konig2001,Jiang2004}.

We report measurements tackling the question of the
coherence of elastic and inelastic cotunneling through QDs~\cite{Averin1990,Franceschi2001}.
Decoherence is generated by which-path detection \cite{Stern1990}.
Inelastic processes are generally believed to lead to
decoherence. An inelastic cotunneling path cannot interfere
with an alternative elastic cotunneling path, because the former
leaves the QD in an excited state thus leaving a trace, which path
the electron took.
We present experimental evidence for phase-coherent AB
oscillations involving elastic and inelastic cotunneling
processes. Our interferometer structure consists of a
QD molecule embedded in an AB ring, similar to
Ref.~\onlinecite{Holleitner2001} thus
realizing systems considered
theoretically~\cite{Loss2000,Konig2001,Jiang2004,Akera1993}.






\begin{figure}[tbhp]
\centering
\includegraphics[width=8.6cm]{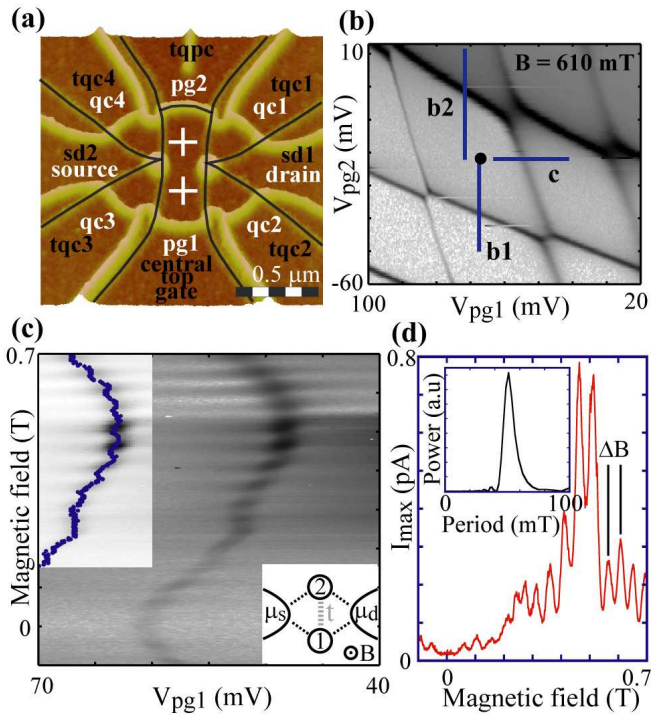}
\caption{\label{figure1} (a) SFM-micrograph of the structure.
In-plane gates (white letters), Titanium oxide lines (black lines)
and top gates (black letters) are indicated. The positions of the
QDs are illustrated by white crosses. (b) Charge stability diagram
of the double QD system. The 9th root of the conductance is
plotted. (c) Conductance peak of dot~1 as a function of magnetic
field and gate pg1 [as indicated by the horizontal line `c' in
(b))]. The upper inset shows the same peak with a line indicating
its maximum. Schematic of the double quantum dot embedded in the
AB-ring in the lower inset (d) The maximum of the conductance peak
as a function of magnetic field. The AB period is about 50~mT
(Fourier analysis shown as inset).}
\end{figure}

The sample shown in Fig.~\ref{figure1}(a) is based on a Ga[Al]As heterostructure with a two-dimensional
electron gas (2DEG) 34~nm below the surface. It was
fabricated by multiple layer local oxidation with a scanning force
microscope (SFM)~\cite{Held1999}: The
2DEG is depleted below the oxide lines written on the
GaAs cap layer. A thin Titanium film is then evaporated on top and
cut by local oxidation into mutually isolated parts acting
as top gates.

The resulting AB-interferometer [Fig.~\ref{figure1}(a)] has a
source and drain opening transmitting at least one mode and being
tunable by the top gates sd1 and 2. One QD is embedded in each arm
of the ring. The two dots are tunnel coupled via a quantum point
contact (QPC) which constitutes an internal connection between the
two branches of the ring. The strength $t$ of this coupling can be
tuned with the central top gate from the tunneling to the open
regime. The two oxide dots forming this constriction by depleting the 2DEG
will be referred to as `antidots' below.
Each QD is coupled to the ring by two QPCs tunable via the top
gates tqc1--4. The in-plane gates pg1 and pg2 are used as plunger
gates for dot~1 and 2, respectively.
Topologically the sample is
similar to those of Refs.~\onlinecite{Holleitner2001} and
\onlinecite{Hatano2004}.

The conductance of the system
was measured in a two-terminal AC lock-in setup at about
80~mK electronic temperature.
For weak interdot coupling with the dots strongly coupled to the
ring the conductance shows an AB period of 22~mT with a visibility (i.e., the ratio of the AB oscillation
amplitude and the magnetic field averaged current) up to 0.2 consistent with
interference around the entire ring. With negative voltages
applied to tqc1--4 the dots can be tuned into the Coulomb blockade
regime.

The leverarms of all gates agree with expectations based on sample
geometry. Each dot has a charging energy of
about 0.7~meV. In our shallow, top-gated structures
it is strongly reduced by image charges in the
top gates. Based on the model calculation in
Ref.~\onlinecite{ihn2003} we find a dot radius of 66~nm, only
slightly larger than that quoted in
Ref.~\onlinecite{Holleitner2001}. The estimated number of
electrons in each dot is about 30. We find single-particle level spacings of the order
of 0.1~meV from nonlinear transport
measurements.

In Fig.~\ref{figure1}(b) the charge stability diagram of the
double dot system is shown with a magnetic field of 610~mT applied
normal to the plane of the 2DEG. It shows the well-known
hexagon pattern formed by regions of constant charge in the two
dots \cite{Sigristnew2004,Hofmann1995}. The 9th root of the
conductance is plotted, enhancing the visibility of the
small cotunneling current. This nonlinear scale is used for all
grayscale figures in this paper, except Fig.~\ref{figure3}(b).
From the offset of conductance peaks at the anticrossings
in Fig.~\ref{figure1}(b) the capacitive interdot coupling is estimated
to be about a tenth of the intradot charging energy and
twice the thermal smearing of conductance resonances.

In Fig.~\ref{figure1}(c) we demonstrate that the field scales for
energy level crossings in the QDs (i.e.,
fluctuations of the conductance peak positions with magnetic
field) and for the AB effect are well separated. To
this end a conductance peak of dot~1 was measured as a function of
$V_\mathrm{pg1}$ and magnetic field while keeping dot~2
off-resonance along the line `c' in
Fig.~\ref{figure1}(b). The peak
shifts smoothly with magnetic field on the scale of a few hundred
mT (about one flux quantum through the dot). The
amplitude of the peak oscillates on a smaller
magnetic field scale with a period $\Delta B\approx 50$~mT.

In order to show the AB-oscillations in more detail we plot in
Fig.~\ref{figure1}(d) the height of the conductance peak as a
function of magnetic field extracted from this measurement [upper
left inset of Fig.~\ref{figure1}(c)] together with its Fourier
transform. For this parameter setting the period $\Delta B$ of the
oscillations corresponds to an area of 165~nm radius, i.e., to
interference paths encircling only one of the two antidots. The
oscillations indicate phase-coherent transmission through both
QDs. The oscillation amplitude is a significant fraction (up to
0.5) of the total current showing that the phase-coherent
contribution to the total current is also significant. This is
quantified by the visibility, the ratio of the AB oscillation
amplitude and the magnetic field averaged current. Visibilities of
up to 0.8 were observed on resonances of dot~1 in some parameter
regions. This is an enormous number if compared to the
visibilities
published in Ref.~\onlinecite{Holleitner2001}.
AB-oscillations with dot~2 on and dot~1 off resonance
were similar, but had a smaller visibility.

Only dot~1 is on resonance in
Fig.~\ref{figure1}(d) while dot~2 allows an elastic
cotunneling current. No AB effect was observed in this regime in
Ref.~\cite{Holleitner2001}. Fig.~\ref{figure1}(c) shows
that in our experiment AB oscillations are even observed
when both dots are in the elastic cotunneling regime,
far away from conductance peaks. In such regions the visibility
can take values of more than 0.8 in
this sample. This value is a conservative estimate accounting for a 4~fA
uncertainty in the offset of the current-voltage converter.

\begin{figure}[tbhp]
\centering
\includegraphics[width=8.6cm]{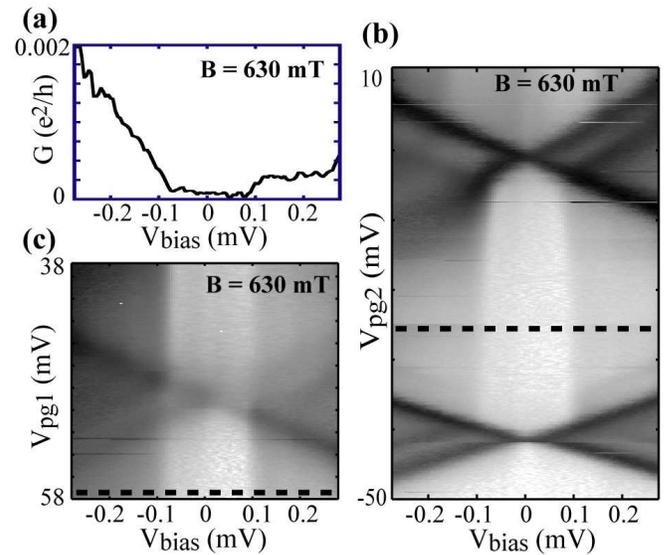}
\caption{\label{figure2} (a) Differential conductance measured as
a function of $V_\mathrm{bias}$ [in-plane gates fixed at black dot
in Fig.~\ref{figure1}(b)]. This curve is indicated in (b) and (c)
and in Fig.~\ref{figure3}(a) as a dashed line. (b) Differential
conductance measured along line (c) in Fig.~\ref{figure1}(b) as a
function of $V_\mathrm{bias}$ and $V_\mathrm{pg1}$. (c)
Differential conductance measured along lines (b1) and (b2) in
Fig.~\ref{figure1}(b) as a function of $V_\mathrm{bias}$ and
$V_\mathrm{pg2}$.}
\end{figure}

We proceed by identifying the inelastic cotunneling onset through
one of the QDs. We have measured Coulomb-blockade diamonds in
the differential conductance shown in Fig.~\ref{figure2}.
Figure~\ref{figure2}(a) shows the differential conductance as a
function of $V_\mathrm{bias}$ taken at a magnetic field of 630~mT
in the center of a hexagon as indicated by the black dot in
Fig.~\ref{figure1}(b). A current step found for positive
$V_\mathrm{bias}$ indicates the onset of inelastic
cotunneling~\cite{Franceschi2001}. Coulomb diamonds for dot~2
[Fig.~\ref{figure2}(b)] measured along the lines `b1' and `b2'
[Fig.~\ref{figure1}(b)] show the typical situation observed in
single dots: the inelastic onset depends on the number of
electrons on dot~2 and is related to excited states outside the
Coulomb-blockaded region~\cite{Franceschi2001}. For dot~1
[Fig.~\ref{figure2}(c)] measured along line `c' in
Fig.~\ref{figure1}(b), a superposition of Coulomb diamonds and an
inelastic cotunneling onset in the current is observed. The
inelastic onset is not affected, if an electron is added to dot~1.
We conclude that depending on bias voltage, the current through
dot~2 is dominated either by elastic or inelastic cotunneling
while the current through dot~1 involves elastic cotunneling only.

As a next step we investigate the phase-coherence of the elastic
and inelastic processes. We explore the
magnetic field dependence of the inelastic cotunneling onset and
look for AB oscillations. To this end both dots are kept in the
cotunneling regime with the in-plane gates fixed [
black dot in
Figure~\ref{figure1}~(b)]. The differential conductance as a function
of magnetic field and $V_\mathrm{bias}$ is shown in
Fig.~\ref{figure3}(a). Two inelastic cotunneling onsets (marked by
arrows) are observed which both depend strongly on magnetic field.
Faint vertical stripes with the period of
interference around one antidot indicate the presence of AB
oscillations across the top right inelastic onset in Fig.~\ref{figure3}(a).

The inelastic onset in the black rectangle measured with higher
resolution is plotted in Fig.~\ref{figure3}(b). The AB
oscillations in the elastic cotunneling regime for small
$V_\mathrm{bias}$ are faint and gradually disappear with
increasing voltage. At the onset of inelastic cotunneling strong
AB oscillations appear, indicating that the inelastic process
does not impair phase coherence.

Cross sections through the data in Fig.~\ref{figure3}(b) taken along the dashed lines
are depicted in Fig.~\ref{figure3}(c). The phase of the
AB oscillations changes by $\pi$ when we cross the inelastic cotunneling onset.
This can also directly be seen in the grayscale plot of Fig.~\ref{figure3}(b).
It confirms that at the inelastic cotunneling onset there is another
transport channel taking over in the coherent transport through dot~2.
The visibility, in particular in the elastic
cotunneling regime, is exceptionally high indicating that dephasing along the
interfering paths is very weak.
\begin{figure}[tbhp]
\centering
\includegraphics[width=8.6cm]{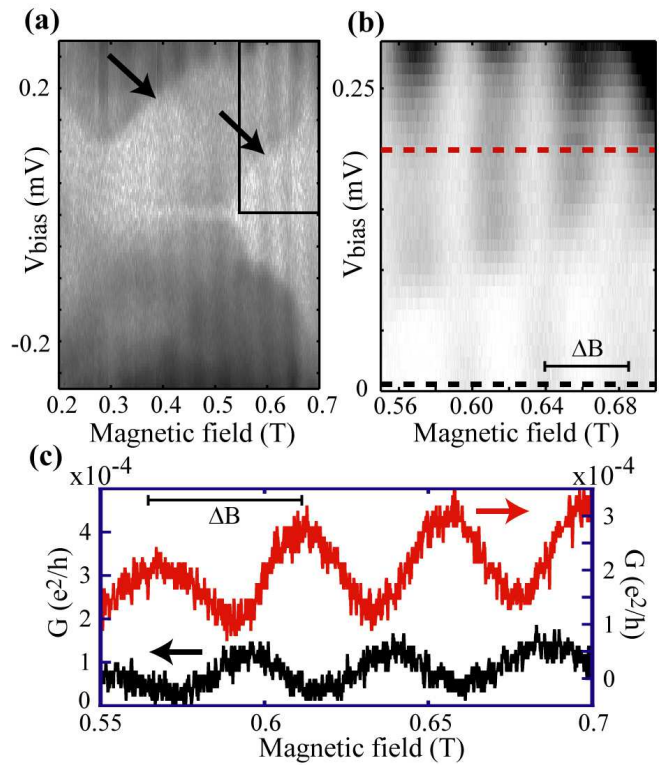}
\caption{\label{figure3} (a) Differential conductance as a
function of magnetic field and $V_\mathrm{bias}$ with both dots in
the cotunneling regime. (b) Detail of (a) inside the black
rectangle. The grayscale is linear. (c) Two traces for small (left
axis) and high bias voltage (right axis) are extracted from
Figure~\ref{figure3}~(b), indicated by dashed lines.}
\end{figure}

AB oscillations from paths around one antidot
were also found in the hexagons surrounding the one for which data was presented above.
The oscillation amplitude depends on the position in the hexagon, being
weakest at the hexagon center and increasing towards the boundaries,
consistent with standard cotunneling models~\cite{Averin1990}.

Similar measurements performed in the regime of weaker tunnel
coupling between the QDs exhibit AB oscillations with a period of
22~mT. The period corresponds to interfering paths encircling both
antidots, i.e., to the whole ring area. In this regime, the
oscillations were only observed in the inelastic cotunneling
regime (visibility about 0.05), because the elastic cotunneling
current was smaller than our current-noise level of about 5~fA.

\begin{figure}[tbhp]
\centering
\includegraphics[width=8.6cm]{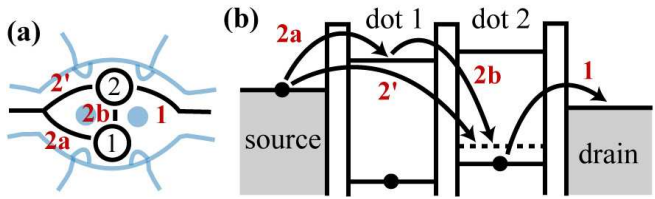}
\caption{\label{figure4} (a) Example for a pair of possible interfering paths
in the AB-interferometer. (b) The same interfering paths in the energy-level
diagram.}
\end{figure}

The observed AB period consistent with paths around one antidot [Fig.~\ref{figure4}(a)]
implies interference between conventional cotunneling through one
dot via one virtual state [e.g., processes 1 and 2' in Fig.~\ref{figure4}(a) and (b)] and cotunneling over at least two
intermediate virtual states in dot~1 and 2 [e.g., processes 1, 2a and 2b in Fig.~\ref{figure4}(a) and (b)].
The processes shown in Fig.~\ref{figure4} are one set out of several
that would lead to the observed interference and we can neither exclude
nor prove that different sets of processes would interfere coherently.
All we can state is that correlated
tunneling of more than one electron is required and not detrimental for the observation of
this kind of interference.

We interpret the phase change between elastic and inelastic
cotunneling observed in Fig.~\ref{figure3}(c) as the fingerprint
of the excited state in dot~2. The relative phase of the
propagating electron between its entrance and exit point contacts
depends on the wave function involved. Our measurement shows that
there is a phase change of $\pi$ when the state involved in
transport changes from the ground- to the excited state. The value
of $\pi$ is compatible with the phase rigidity expected for a
two-terminal measurement.

The huge numbers found for the visibilities in our experiment are
remarkable. We argue that the involved cotunneling processes
require a short tunneling time of the order of $h/U\sim 10$~ps
($U$ is half the charging energy) which is short
compared to dephasing times of more than $1$~ns reported in other
experiments ~\cite{Yacoby1995}. Perhaps even higher order
cotunneling processes than those mentioned above as examples, can
take place.

Why do we measure no significant suppression of the AB
interference by inelastic cotunneling? Considering the data, the
most likely explanation is that exemplified in Fig.~\ref{figure4}
where the excited state in dot~2 does not allow which path
detection. This is conceivable, if the two interfering paths both
start in the source contact and end in dot~2, one taking the
detour via dot~1. A second possible scenario would require that
the excited state extends into both dots and does therefore not
allow which-path detection~\cite{Loss2000}. It is conceivable that
other situations exist which combine inelastic tunneling processes
with phase coherence of the entire system.


Our experiment is a significant step towards the proposed
detection of entanglement via the AB effect~\cite{Loss2000}.
Beyond the demonstration of coherence in the elastic and inelastic
cotunneling regime we have chosen the hexagon investigated above in such
a way that it is bounded by states which move in a highly
correlated fashion with magnetic field. Such states are commonly
believed to be spin-pairs~\cite{Luscher00}, i.e., states of different spin but with
the same orbital wave function. We therefore speculate that in
each dot one unpaired spin occupies the highest orbital level.
However, the exchange coupling necessary for the formation of
singlet and triplet states was probably too low in our experiment
due to the moderate tunnel coupling between the dots.

In conclusion, we have experimentally demonstrated the
phase coherence of inelastic tunneling in the elastic and inelastic cotunneling regimes in
quantum dots. Visibilities of more than 0.8 were measured
indicating that the phase-coherent current dominates the
conductance. A phase jump of $\pi$ was detected at the onset of
inelastic cotunneling processes. We anticipate that cotunneling
processes could be employed in applications where a huge degree of
phase coherence is crucial.

\begin{acknowledgments}
We thank R. Schleser for valuable discussions. Financial support
by the NCCR Nanoscience through the Swiss Science Foundation
(Schweizerischer Nationalfonds) is gratefully acknowledged.
\end{acknowledgments}



\end{document}